\documentclass{article}

\usepackage{times}
\usepackage{graphicx} 
\usepackage{subfigure} 
\usepackage{natbib}
\usepackage{algorithm}
\usepackage{algorithmic}

\usepackage{hyperref}

\usepackage[accepted]{icml2017}

\icmltitlerunning
{RNN-based Early Cyber-Attack Detection for the Tennessee Eastman Process}

\begin{document} 

\twocolumn[
\icmltitle
{RNN-based Early Cyber-Attack Detection for the Tennessee Eastman Process}



\icmlsetsymbol{equal}{*}

\begin{icmlauthorlist}
\icmlauthor{Pavel Filonov}{KL}
\icmlauthor{Fedor Kitashov}{KL}
\icmlauthor{Andrey Lavrentyev}{KL}
\end{icmlauthorlist}

\icmlaffiliation{KL}{Kaspersky Lab, Moscow, Russian Federation}

\icmlcorrespondingauthor{Andrey Lavrentyev}{Andrey.Lavrentyev@kaspersky.com}

\icmlkeywords{Tennessee Eastman Process, multivariate time series, anomaly early detection,  RNN, forecasting model, NAB-metric}
\vskip 0.3in
]




\printAffiliationsAndNotice{}

\begin{abstract} 

An RNN-based forecasting approach is used to early detect anomalies in industrial multivariate time series data from a simulated Tennessee Eastman Process (TEP) with many cyber-attacks. This work continues a previously proposed LSTM-based approach to the fault detection in simpler data. It is considered necessary to adapt the RNN network to deal with data containing stochastic, stationary, transitive and a rich variety of anomalous behaviours. There is particular focus on early detection with special NAB-metric. A comparison with the DPCA approach is provided. The generated data set is made publicly available. 

\end{abstract}

\section{Introduction}

Modern Industrial Control Systems (ICS) deals with multivariate time series data of technological processes: sensors and controls signals. Comprising a cyber components, ICSs are a target of cyber-attacks (for example ~\cite{Steel14}), that can modify sensor and controls values, or the parameters of control logic (set points). Such cyber-attacks can be detected as an anomalies in technological signals. This raises the issue  of early anomaly detection. 

Different approaches have been proposed to detect anomalies in industrial data.  Anomalies can arise for different reasons, besides cyber-attacks:  equipment malfunctions, human errors, analogous signals interruptions, etc. Here we provide only a short overview of such approaches:  RNN-based  ~\citep{Aircraft16}, LSTM-based forecasting ~\citep{KL16,Malhotra15} and encoder-decoder ~\cite{Malhotra16}, clustering based  ~\cite{TEP_DoS_15},  PCA, DPCA, FDA, DFDA, CVA, PLS  ~\citep{FDDinIS}, one-class SVM and segmentation ~\cite{Petrol15}, change point detection ~\cite{ChPoint13}, process invariants ~\cite{ProcInv16}.
 
One of the main problems with the verification of proposed approaches is the lack of available industrial datasets with labelling of normal and anomalous behaviour as well as the absence of rich anomalous behaviour examples. Finding data from real objects under cyber-attacks is problematic because these are quite unique incidences and industry vendors do not want to share such data. Experimenting with attacks on real {\it test} objects is not a solution because  it is very costly. One of a possibility for generating anomalous behaviour is data augmentation as in ~\citep{Augm16}. Another possibility is to use a mathematical model of a cyber-physical system for both physics and control dynamics and simulate multiple  realistic cyber-attacks. In our previous work ~\citep{KL16} we used this approach with a gasoil heating loop process (GHL) ~\cite{GHL48} implemented with the Modelica tool. The generated data is quite rich but it lacks of some stochastic properties and reflects a rather simple control logic. 

In the current paper we use the well-known TEP model ~\citep{TE_DownsVogel_93,TE_Ricker} which allows rich and realistic datasets to be generated. Cyber-attack simulation using TEP was proposed in ~\citep{TE14_Krotofil} and  implemented in the Matlab/Simulink tool and .NET code.  We used our own implementation of the TEP model completely in Python code which allowed us to simulate a lot of cyber-attacks  and generate datasets as well as a streaming data.  

To detect anomalies in TEP data we further developed the RNN-based forecasting approach that we used for GHL data. TEP data requires the RNN network to be adapted in order to deal with stochasticity, stationary and transitive behaviours.  We also focused more on early detection and for this purpose used Numenta Anomaly Benchmark (NAB) metric ~\citep{NAB}. We provide a comparison with the fault detection approach traditionally used for TEP based on DPCA  ~\citep{FDDinIS}, and which we combined here with the NAB-metric.

\section{Dataset Description}
\label{sec:data_desc}

\begin{figure}[ht]
\vskip 0.2in
\begin{center}
\centerline{\includegraphics[width=\columnwidth]{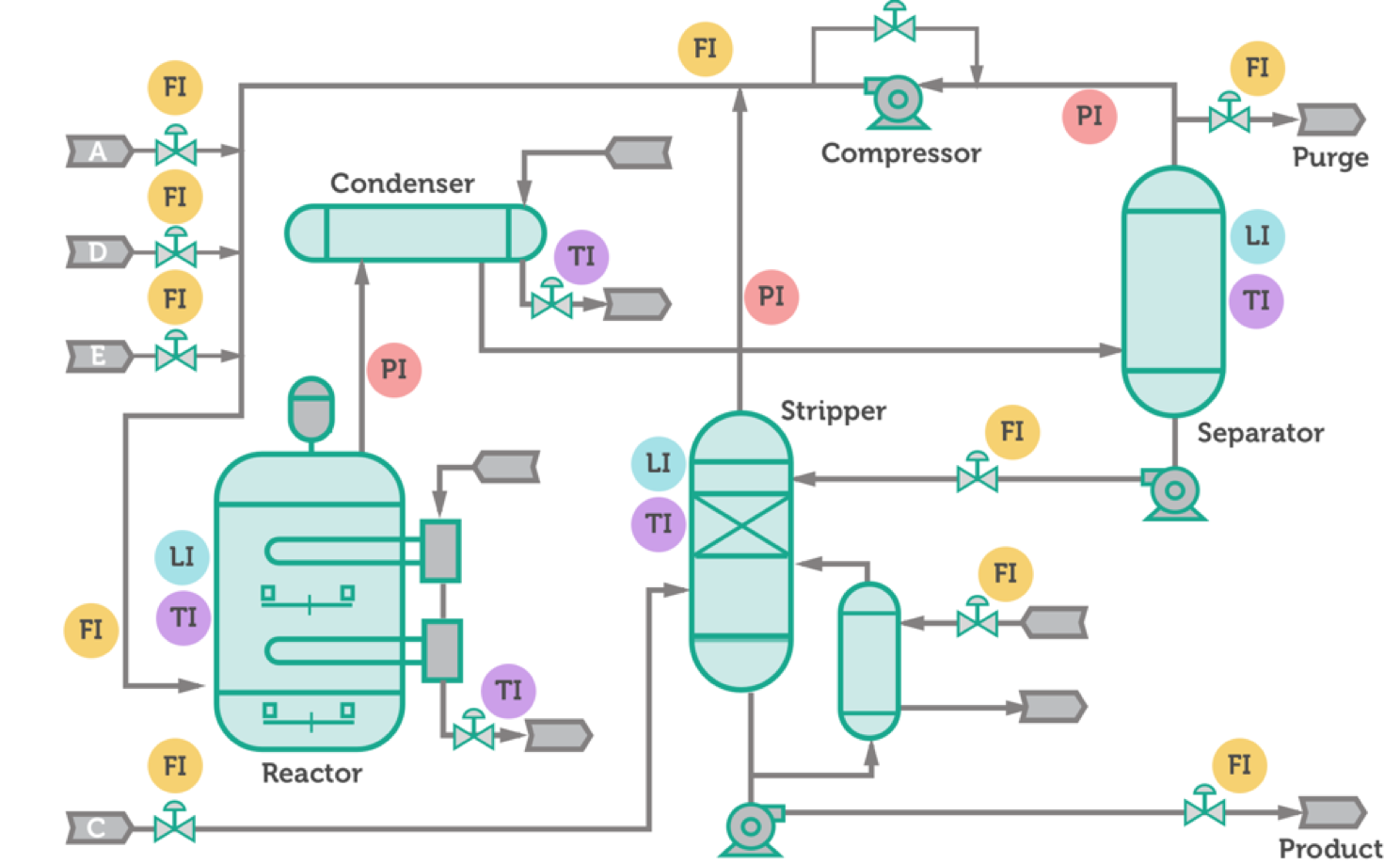}}
\caption{Tennessee Eastman Process}
\label{fig:TEP}
\end{center}
\vskip -0.2in
\end{figure}

The TEP model is represented in Figure ~\ref{fig:TEP}.
\noindent
It was simulated at different normal modes and under cyber-attacks.  The generated datasets characteristics are represented in Table \ref{tab:dataset}.

\begin{table}[h]
\vskip 0.15in
\begin{center}
\begin{small}
\begin{tabular}{|l|l|}
\hline
\abovespace\belowspace
\bf 59 & \bf Time Series Dimension  \\
\hline
$41$  &  sensors (MEAS - measurements)  \\
$12$  &  controls (MV - manipulated variables)  \\
$1$   &  MEAS attack indicator  \\
$1$   &  MV attack indicator \\
$1$   &  SP (set point) attack indicator  \\
$3$   & special variables (state, product rate, hourly cost)\\
\hline
\abovespace\belowspace
\bf &  \bf Plant Modes  \\
\hline
$7$  & normal modes  \\
$28$ & transient modes \\
\hline
\abovespace\belowspace
 &  \bf Attack Types\\
\hline
 &  DoS (value is frozen) \\
 &  Integrity (value is changed) \\
 & Noise (value + noise) \\
\hline
\abovespace\belowspace
 &  \bf Attack Series (\#, Type, MV/MEAS/SP,  duration) \\
\hline
 &  \#21: Integrity:  MEAS  "reactor temperature", \\
 & \hspace{16pt}    0.012-0.027 h\\
 &  \#22: DoS: MV "Stripper liquid product flow", \\ 
 & \hspace{16pt} MEAS "Stripper level", \\ 
 & \hspace{16pt} MEAS "Stripper underfow", 5.663-25.019 h\\
 &  \#23: DoS: MV "D feed flow", 10 h\\
 &  \#24: Noise:  MV "C feed flow", MV "Purge flow", \\
 & \hspace{16pt} MEAS "Stripper underfow",\\ 
 & \hspace{16pt} MV "Stripper steam flow", 7.727 - 71.291 h\\
\hline
\abovespace\belowspace
$1000$ & \bf Points per Hour  \\
\hline
\abovespace\belowspace
\ &  \bf Training set (duration hours) \\
\hline
$201$  & samples with one normal mode ($120$ h) \\
$336$ & samples with transient mode ($120$ h)		\\
\hline
\abovespace\belowspace
 &  \bf Test set (duration hours) \\
\hline 
$142$ & samples with attacks ( $\le 120$ h, till broken) \\
\hline
\end{tabular}
\end{small}
\end{center}
\vskip -0.1in
\caption{TEP dataset characteristics}
\label{tab:dataset}
\end{table}
\noindent
We generated a training dataset  with 201 single-mode and 336 transient-mode samples and a test dataset with 142 MEAS/MV/SP attacks  samples ~\cite{TEP59}. Each sample is a multivariate time series of dimension 59. Besides samples for $7$ single modes of TEP operation we generated samples for 28 transient modes via 4 variants of SP changes for each single mode: decreasing by $2\%$ catalyst $C$ purge, changing product mix by $10\%$,  decreasing product rate by $15\%$, decreasing  reactor pressure by $1-2\%$.  Indicators of attacks in the test dataset  are equal to $1.0$ at the intervals of corresponding attacks (to MEAS, MV or SP). There were three  kinds of attacks used at the MEAS and MV: a) Integrity:  changing a value to something different from that simulated by the TEP-model, b) DoS (denial  of service):  at some point a value of a variable is frozen for the duration of an attack, c) Noise: add nose to value.

An attack on an industrial plant can very quickly reach a critical situation where further model simulation becomes impossible and the plant operation must be stopped.  In order not to make the task of detection too simple, we tuned the attack intervals so that the plant could return to a level of stable operation after an attack, and proposed four series of attacks.

\section{RNN-based Anomaly Detection}

We use RNN-based forecasting model.  Anomaly detection is made on the base of MSE (mean square error) between prediction and observation. 

\subsection{Pre- and Post- Processing}

Input data is normalized (parameters are calculates based on the training dataset).

Prediction square error is summarised and smoothed with EMWA. Smoothing factor $\alpha$ is calculated using the size of input window $w$ as $\alpha = 1-  \exp{(- \frac{\ln{2}}{w})}$.   

The minimal detection threshold value is calculated as $0.999$ quantile from the smoothed error in the training dataset.

\subsection{RNN Architecture and Training}

To cope with the TEP dataset, we adopted the previously used LSTM architecture for the GHL dataset in a way that is represented in Table \ref{tab:arch}.

\begin{table}[h]
\vskip 0.15in
\begin{center}
\begin{small}
\begin{tabular}{|l|c|c|c|c|c|c|}
\hline
\abovespace\belowspace
\bf Dataset & \bf Cell & \bf Layer & \bf Memory & \bf Dropout & \bf Window\\
\hline
\abovespace
GHL  &  LSTM &  2x64 & stateful & $0.1$  & $120$ \\
\belowspace
TEP  & GRU &  2x64 &  stateless & no   & $100$ \\ 
\hline
\end{tabular}
\end{small}
\end{center}
\vskip -0.1in
\caption{RNN architecture for GHL and TEP datasets}
\label{tab:arch}
\end{table}

For both datasets we use  stacked RNN with 2 hidden layers, each with 64 cells. The input window is equal to the prediction window.  ReLU as an activation function for hidden layers and linear activation function for the output layer are used.

\begin{figure}[ht]
\vskip 0.1in
\begin{center}
\centerline{\includegraphics[width=\columnwidth]{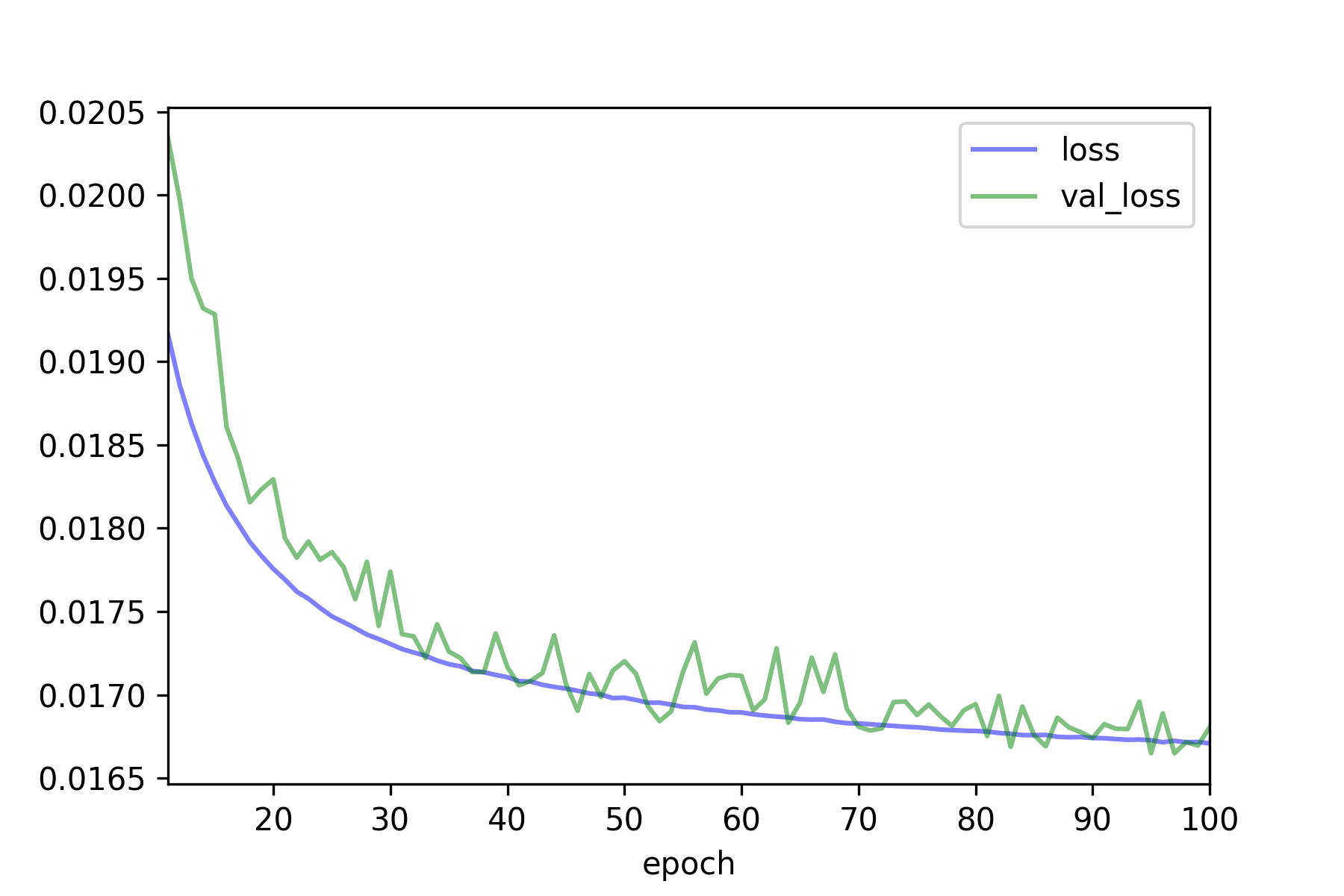}}
\caption{Loss-function value vs epoch number for training (loss) and validation (val-loss) datasets}
\label{fig:loss}
\end{center}
\vskip -0.1in
\end{figure}

To train RNN we use MSE  loss-function and the RMSProp algorithm. Learning step equals $0.001$. Number of epochs equals $100$. Average time of one  training epoch is  $~ 70$ seconds with batch size = 2048 and hardware Tesla P40, Intel Xeon CPU E5-2650 v4  2.20GHz. The resulting dependency of loss-functions vs epoch for training and validation datasets is represented in Figure \ref{fig:loss}.

\begin{figure}[ht]
\vskip 0.1in
\begin{center}
\centerline{\includegraphics[width=\columnwidth]{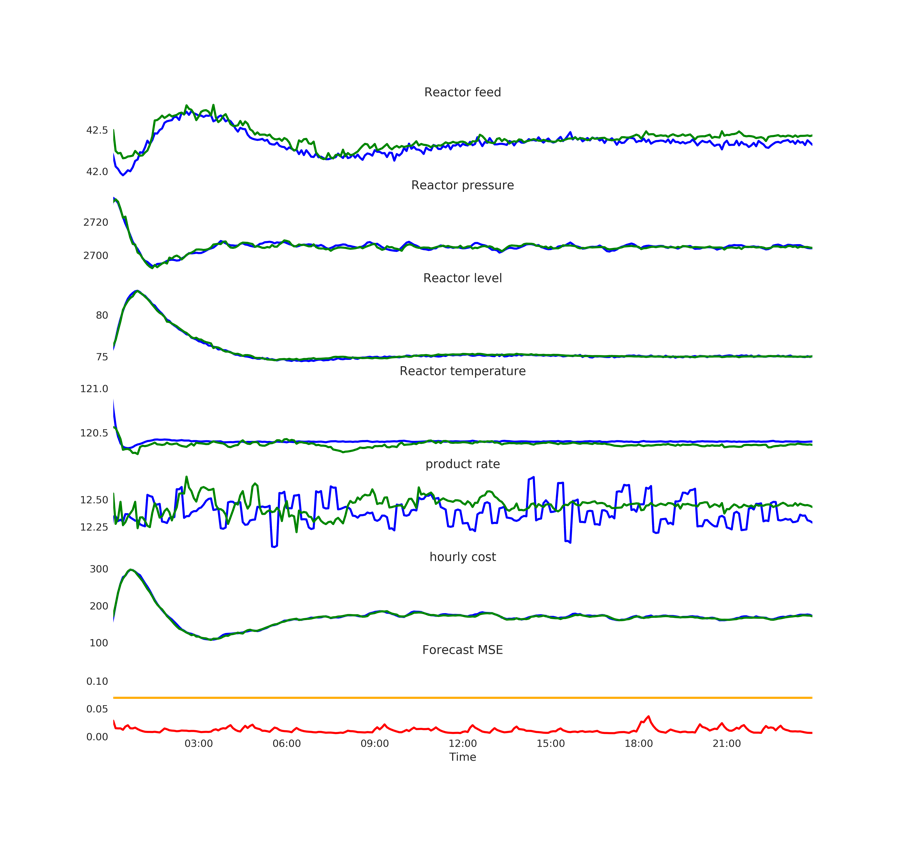}}
\caption{Example of RNN prediction (green) for a single mode normal behaviour sample}
\label{fig:single-mode}
\end{center}
\vskip -0.1in
\end{figure}

\begin{figure}[ht]
\vskip 0.2in
\begin{center}
\centerline{\includegraphics[width=\columnwidth]{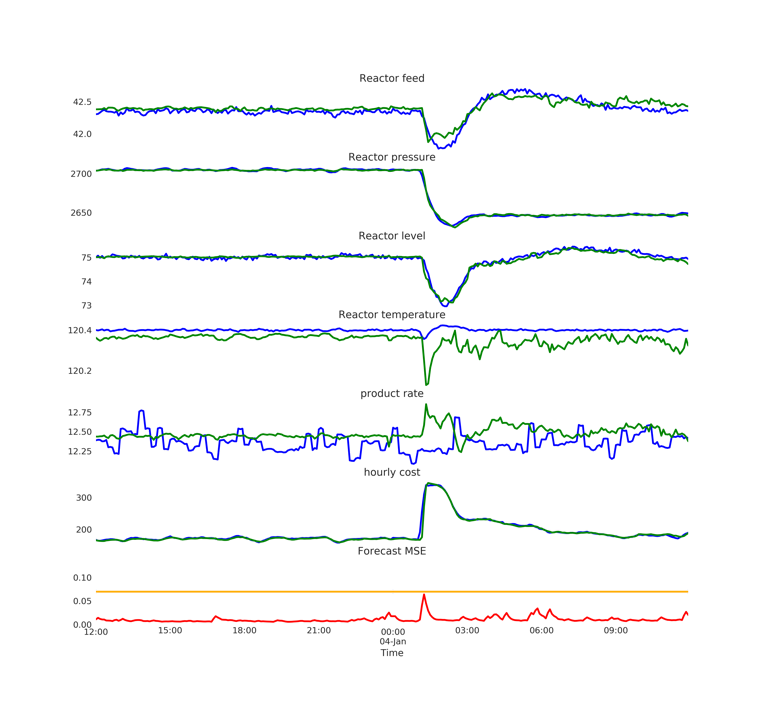}}
\caption{ Example of RNN prediction (green) for a transient mode normal behaviour sample}
\label{fig:transient-mode}
\end{center}
\vskip -0.2in
\end{figure}

\begin{figure}[ht]
\vskip 0.2in
\begin{center}
\centerline{\includegraphics[width=\columnwidth]{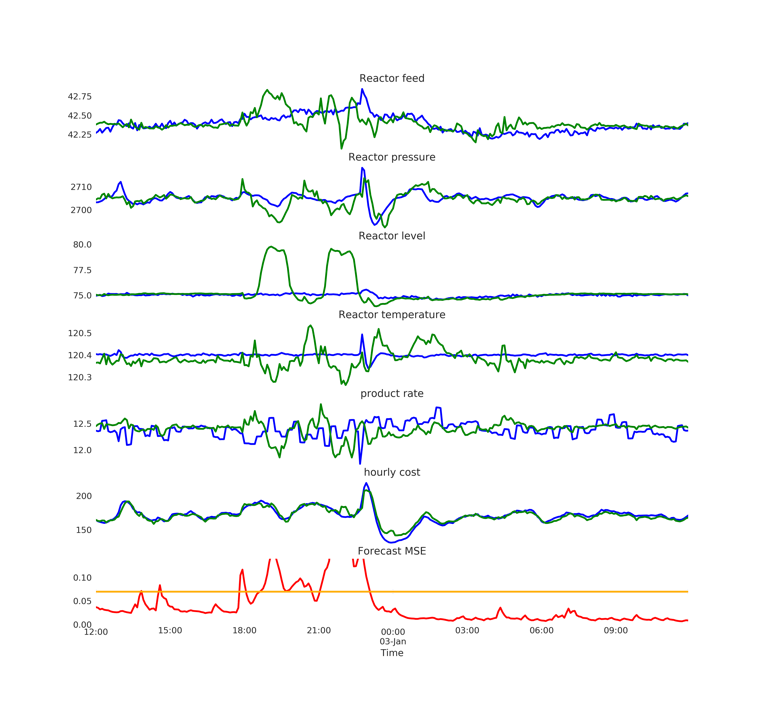}}
\caption{Example of RNN prediction (green) for an MEAS-attack sample }
\label{fig:meas-attack}
\end{center}
\vskip -0.2in
\end{figure}

Examples of trained RNN model prediction for  a single mode normal behaviour sample is represented in Figure \ref{fig:single-mode}, for a transient mode sample in Figure \ref{fig:transient-mode},  and for an MEAS attack sample in Figure \ref{fig:meas-attack}.

\subsection{Quality Metric}

To compare the results of different anomaly detection approaches we selected the NAB-metric that scores in range  $s \in [-1.0, 1.0]$ ($s=1.0$ if detection is at the anomaly beginning,  $s=0.0$ if detection is at the end of anomaly window, $s \in (-1.0,0.0)$ if detection is not too far from the end of anomaly window, $s=-1.0$ otherwise). Table~\ref{tab:nab-weights} shows standard profile weights ~\cite{NAB} for TP, TN, FP, FN for the NAB-metric. 
\begin{table}[h]
    \centering
    \begin{tabular}{|l|l|l|}
    \hline
        & \bf Positive & \bf Negative  \\
    \hline
    \bf True & $1.0$ & $1.0$   \\
    \bf False & $0.11$ & $1.0$   \\
     \hline
    \end{tabular}
    \caption{Standard profile weights for the NAB-metric}
    \label{tab:nab-weights}
\end{table}

Experimenting with different kinds of attacks on the TEP we observed that the anomaly window is not necessary equal to the attack interval. Quite often the consequence of an attack, which is also anomalous behaviour, continues after the attack has stopped. So, selecting a correct anomaly window for the NAB metric is quite a tricky process. To average this out we use an anomaly window equal to twice the attack interval.

The RNN-based detector was tested under different detection thresholds. Several cyber-attacks datasets were concatenated in one.

\subsection{Comparison with DPCA}

Working with the GHL dataset we found that the most successful alternative to the LSTM-based approach is PCA. Here we compare our RNN-based approach with dynamic PCA (DPCA).  

DPCA parameters are:
time window size - 10; 
space dimension -  590;
number of main components - 19 (Kaiser rule $\lambda > 1.0$).

With DPCA we were only able  to train separate models for each TEP single operation mode. For transient mode we faced with many false positives (FP) detection with DPCA. So, we ignored that cases and calculated scores for DPCA as an average of the scores for each single mode ($m$): $$\overline{\textrm{DPCA}} = \sum_{m = 0}^{6}  \textrm{DPCA}_{(m)}$$

We tested RNN and DPCA  on the TEP dataset using the NAB-metric. Anomalies detections results are shown in Table 
 \ref{tab:comparison}.

\begin{table}[h]
    \centering
    \begin{tabular}{|l|r|}
    \hline
     \bf Method (attacks series) & \bf NAB-score \\
    \hline
     Ideal detector & $1.000$\\
     RNN (all) & $0.373$\\
      $\overline{\textrm{DPCA}}$  (all)  & $0.086$ \\
     RNN (except \#23) & \bf $0.803$\\
     $\overline{\textrm{DPCA}}$ (except \#23) & $0.649$ \\
     \hline
    \end{tabular}
    \caption{ RNN vs DPCA NAB-scores}
    \label{tab:comparison}
\end{table}

We connect the decrease in the RNN and DPCA detection score in the NAB-metric for attacks on MV \#23 ("D-feed flow DoS") with the TEP physics, i.e. the consequences of control changes taking place for quite a long time after an attack.

\section{Conclusion}
\label{seq:conc}

The RNN-based approach with GRU stateless cells and without dropout is capable of effectively dealing with stochasticity, stationarity, transient and anomalous behaviour in a realistic TEP dataset. The NAB-metric makes it possible to validate the model for early detection. A comparison with DPCA shows that the RNN-based approach has better scores for MEAS and SP attacks.  Attacks on MV are detected with RNN with some delay, which we explain by the longer anomaly window of the consequences of such attacks. We  also found that DPCA model can be trained only for a separate single mode,  and for a transient mode DPCA gives many false positives (FP). From a practical point of view of industrial anomaly detection application,  it is more convenient to have one trained model for all kinds of plant modes, what we achieved only with RNN approach. 

The generated TEP datasets with normal and anomalous behaviour caused by cyber-attacks are made publicly available.

\section*{Acknowledgements} 
The authors are sincerely grateful to Konstantin Kiselev for the implementation of the TEP model on Python and Artem Vorontsov for useful discussions.

This work was supported by the Kaspersky Lab.

\bibliography{icml2017_tsw}
\bibliographystyle{icml2017}

\end{document}